\documentclass[showpacs,preprintnumbers,showkeys,aps]{revtex4}
\usepackage{graphicx, epsfig, epstopdf}
\usepackage{amsmath, amsfonts, amssymb}
\usepackage{bm, array}
\usepackage[usenames]{color}
\usepackage{hyperref}
\usepackage{eurosym}
\usepackage{mdframed}
\usepackage{grffile}
\usepackage{amssymb}
\usepackage{caption}

\def\be{\begin{equation}}
\def\ee{\end{equation}}
\def\beq{\begin{eqnarray}}
\def\eeq{\end{eqnarray}}

\begin{document}

\title{Rotating Thin-Shell Wormhole}
\author{A. Ovgun}
\email{ali.ovgun@emu.edu.tr}
\affiliation{Physics Department, Eastern Mediterranean University, Famagusta, Northern
Cyprus, Mersin 10, Turkey}
\date{\today}

\begin{abstract}
We construct a rotating thin-shell wormhole using a
Myers-Perry black hole in five dimensions, using the Darmois-Israel junction conditions. The stability of the wormhole is analyzed under perturbations. We find that exotic matter is required at the throat of the wormhole to keep it stable. Our analysis shows that stability of the rotating thin-shell wormhole is possible if suitable parameter values are chosen.
\end{abstract}

\pacs{04.50.Gh,04.20.Gz,04.20.-q }
\keywords{Wormholes; Thin-Shells; Stability; Darmois-Israel formalism;
4+1-dimensions; Myers-Perry Black Hole}
\maketitle

\section{Introduction}

One of the most challenging problem in Einstein's general relativity finding stable wormholes with a minimum amount of exotic matter or with completely normal matter \cite{Mor,V1,V2}. In this regard we note that rotating wormholes present
more alternatives because of their extra degrees of freedom. Calculations in
rotating spacetimes are clearly difficult to apply to stability analyses of
wormholes. Moreover, there are only a few works on the rotating thin-shell
wormholes (RTSWs) built in 2+1 dimensions and in 3+1 dimensions with some
approximations \cite{lemos,sak1}. On the other hand, there are many
papers demonstrating the construction of thin-shell wormholes using different modified theories or extra/lower dimensions \cite%
{visser2,visser3,eiro,ali1,richarte,richarte2,ali2,vare,rahaman,rahaman2,baner1,eiro2,eid,cam,sh1,sh2,ric1,ric2,eiro3,eiro4}. Recently, a rotating thin-shell wormhole was constructed and its
thermodynamics worked in 2+1 dimensions, where the problem is more
malleable than in four or more dimensions \cite{lemos}.

This paper constructs a rotating thin-shell wormhole in five
dimensions by employing Visser's cut-and-paste technique. For the first time in the literature, we address the stability of a five-dimensional (5-d) rotating
thin-shell wormhole. In our analysis, we use the five-dimensional rotating
Myers-Perry black hole solution with all angular momenta equal, which has been used previously to work on collapsing thin-shells with rotation \cite{delsate}.

\section{Constructing The Rotating Thin-Shell Wormhole}

The 5-d rotating Myers-Perry (5DRMP) black hole solution, which is the
generalization of the Kerr solution to higher dimensions, is given by the following space-time metric \cite{myers,kunduri}:
\begin{equation}
ds^{2} =-F(r)^{2}dt^{2}+G(r)^{2}dr^{2}+r^{2}\widehat{g}_{ab}dx^{a}dx^{b} 
+\,H(r)^{2}\left[ d\psi +B_{a}dx^{a}-K(r)dt\right] ^{2}\,,
\label{eq:metric}
\end{equation}%
in which 
\begin{eqnarray}
G(r)^{2} &=&\left( 1+\frac{r^{2}}{\ell ^{2}}-\frac{2M\Xi }{r^{2}}+\frac{%
2Ma^{2}}{r^{4}}\right) ^{-1}\,,  \label{eq:metricfuncs1} \\
H(r)^{2} &=&r^{2}\left( 1+\frac{2Ma^{2}}{r^{4}}\right) \,,\qquad K(r)=\frac{%
2Ma}{r^{2}H(r)^{2}}\,,  \label{eq:metricfuncs2} \\
F(r) &=&\frac{r}{G(r)H(r)}\,,\qquad \Xi =1-\frac{a^{2}}{\ell ^{2}}\,,
\label{eq:metricfuncs3}
\end{eqnarray}%
where $B=B_{a}dx^{a}$ and 
\begin{equation}
\widehat{g}_{ab}dx^{a}dx^{b}=\frac{1}{4}\left( d\theta ^{2}+\sin ^{2}\theta
\,d\phi ^{2}\right) ,\;\;B=\frac{1}{2}\cos \theta \,d\phi \,.
\end{equation}

Note that taking the limit of the Anti-de-Sitter (AdS) length $\ell \rightarrow \infty $ , the
asymptotically flat case can be recovered.
The event horizon is located at the largest real root of $G^{-2}$. One
writes the mass $\mathcal{M}$ and angular momentum $\mathcal{J}$ of the
spacetime as \cite{kunduri}

\begin{equation}
\mathcal{M}=\frac{\pi M}{4}\left( 3+\frac{a^{2}}{\ell ^{2}}\right) \,,\qquad 
\mathcal{J}=\pi Ma\,.
\end{equation}
For convenience we move to a comoving frame to eliminate cross terms in the
induced metrics by introducing \cite{mann3} 
\begin{equation}
d\psi \longrightarrow d\psi ^{\prime }+K_{\pm }(\mathcal{R}(t))dt\,.
\end{equation}%

We choose a radius $\mathcal{R(}t)$,  which is the throat of the wormhole, and take two copies of this manifold \ $%
\tilde{M}_{\pm }$ for the interior and exterior regions with $r\geq \mathcal{%
R}$ to paste them at an identical hypersurface $\Sigma =\{x^{\mu }:t=%
\mathcal{T}(\tau ),\,r=\mathcal{R}(\tau )\}$, which is parameterized by
coordinates $y^{i}=\{\tau ,\psi ,\theta ,\phi \}$ on the 4-d surface.

The line element in the interior and the exterior sides become 
\begin{eqnarray}
ds_{\pm }^{2} &=&-F_{\pm }(r)^{2}dt^{2}+G_{\pm }(r)^{2}dr^{2}+r^{2}d\Omega 
+H_{\pm }(r)^{2}\{ d\psi ^{\prime }+B_{a}dx^{a}+[K_{\pm }(\mathcal{R}%
(t))-K_{\pm }(r)]dt\}^{2}.   \label{eq:comoving}
\end{eqnarray}
For simplicity in the comoving frame, we drop the prime on $\psi ^{\prime }$.
The geodesically complete manifold is satisfied as $\ \tilde{M}=\tilde{M}_{+}$ $U$ $%
\tilde{M}_{-}$. We use the Darmois-Israel formalism to construct the rotating
thin-shell wormhole \cite{israel}. The throat of this wormhole is located at
the hypersurface of $\Sigma $ and to satisfy the Israel junction
conditions, we first define \begin{equation} F_{+}(\mathcal{R})=F_{-}(\mathcal{R})=F(%
\mathcal{R}) \end{equation} and \begin{equation} -F_{\pm }(\mathcal{R})^{2}\left( \mathcal{\dot{T}}%
\right) ^{2}+G_{\pm }(\mathcal{R})^{2}\left( \mathcal{\dot{R}}\right)
^{2}=-1, \end{equation} where dot stands for $d/d\tau$. The extrinsic curvature\ is
calculated from  \begin{equation} k_{\mu \nu }=(g_{\mu \sigma }-n_{\mu }n_{\sigma })\nabla
^{\sigma }n_{\nu }, \end{equation} where the normal vector is \begin{equation} n_{\mu }=F(r)G(r)\left( -%
\mathcal{\dot{R}},\mathcal{\dot{T}},0,0,0\right). \end{equation}
The second junction condition implies%
\begin{equation}
\left[ K_{i}^{j}\right] -\left[ K\right] \delta _{i}^{j}=-8\pi GS_{i}^{j}
\end{equation}%
in which a bracket $\left[ {\ }\right] $ is defined as%
\begin{equation}
\left[ A\right] =A_{\left( o\right) }-A_{\left( i\right) }
\end{equation}%
and the extrinsic curvature tensor is
\begin{equation}
K_{ij}=-n_{\gamma }\left( \frac{\partial ^{2}x^{\gamma }}{\partial
x^{i}\partial x^{j}}+\Gamma _{\alpha \beta }^{\gamma }\frac{\partial
x^{\alpha }}{\partial x^{i}}\frac{\partial x^{\beta }}{\partial x^{j}}%
\right).
\end{equation}%

Next, Einstein's equations on the shell second junction condition are used to obtain the surface energy-momentum tensor of throat chosen as a perfect fluid, as in \cite{del,cola}:
\begin{equation}
\mathcal{S}_{ij}=(\rho +P)u_{i}u_{j}+P\,{\mathfrak{g}}_{ij}+2\varphi
\,u_{(i}\xi _{j)}+\Delta P\,\mathcal{R}^{2}\widehat{g}_{ij},
\label{eq:stress}
\end{equation}%
where $u=\partial _{\tau }$ is the fluid four-velocity and ${\mathfrak{g}}_{ij}$ stands for the induced metric on $\Sigma$. Note that $\xi =H(%
\mathcal{R})^{-1}\partial _{\psi }$, $\widehat{g}_{ij}dy^{i}dy^{j}=\widehat{g%
}_{ab}dx^{a}dx^{b}$ \cite{Herr}, and a perfect fluid~is obtained for $\Delta
P=\varphi =0$ .  Using the Israel junction conditions \cite{israel}, the Einstein's equations for the wormhole produce

\begin{eqnarray}
\rho &=&-\frac{\beta (\mathcal{R}^{2}H)^{\prime }}{4\pi \,\mathcal{R}^{3}}%
\,,\;\varphi =-\frac{\mathcal{J}(\mathcal{R}H)^{\prime }}{2\pi ^{2}\mathcal{R%
}^{4}H}\,,  \label{eq:stress_components} \\
P &=&\frac{H}{4\pi \mathcal{R}^{3}}\left[ \mathcal{R}^{2}\beta \right]
^{\prime },\;\Delta P=\frac{\beta }{4\pi }\left[ \frac{H}{\mathcal{R}}\right]
^{\prime },  \label{eq:stress_components2}
\end{eqnarray}
where primes stand for $d/d\mathcal{R}$ and \begin{equation} \beta \equiv F(\mathcal{R})%
\sqrt{1+G(\mathcal{R})^{2}\mathcal{\dot{R}}^{2}}\,. \label{ddot}\end{equation}

 Without rotation or in the case of a corotating frame, the momentum $\varphi $ and the
anisotropic pressure term $\Delta P$ are equal to zero. Consequently, the static energy and pressure densities at the throat of wormhole $\mathcal{R}$ = $%
\mathcal{R}_{0}$ are given by%
\begin{eqnarray}
\rho _{0} &=&-\frac{F(\mathcal{R}_{0}^{2}H)^{\prime }}{4\pi \,\mathcal{R}%
_{0}^{3}}\,,\;\varphi _{0}=-\frac{\mathcal{J}(\mathcal{R}_{0}H)^{\prime }}{%
2\pi ^{2}\mathcal{R}_{0}^{4}H}\,,  \label{stres0} \\
P_{0} &=&\frac{H}{4\pi \mathcal{R}_{0}^{3}}\left[ \mathcal{R}_{0}^{2}F\right]
^{\prime },\;\Delta P_{0}=\frac{F}{4\pi }\left[ \frac{H}{\mathcal{R}_{0}}%
\right] ^{\prime }.  \label{p0}
\end{eqnarray}
To check the stability of the wormhole, we use the linear equation of state (EoS):
\begin{equation}
P=\omega \rho . \label{eos}
\end{equation}%
Using the Eqns. (\ref{eq:stress_components}), (\ref{eq:stress_components2}), and (\ref{eos}), we obtain the $\beta $ as follows:

\begin{equation}
\beta =-\frac{m_{0}^{1+3\omega /2}}{\mathcal{R}^{2(1+\omega )}H^{\omega }}\,,\label{betaa}
\end{equation}
where $m_{0}$ is a positive constant with dimensions of mass. Furthermore, the dynamics of the throat are described by the thin-shell
equation of motion, which can be obtained using Eqns. (\ref{ddot}) and (\ref{betaa})  as follows: 
\begin{equation}
\mathcal{\dot{R}}^{2}+V_{eff}=0,
\end{equation}%
where the effective potential (for simplicity we choose $l=M=m_{0}=1$) in a static configuration at the throat of the wormhole is calculated as

\begin{equation}
V_{eff}={\frac {-{F}^{2}{\mathcal{R}%
_{0}}^{4}+{\mathcal{R}%
_{0}}^{-4\,\omega}{H}^{-2\,\omega}}{{F}
^{2}{G}^{2}{\mathcal{R}%
_{0}}^{4}}}.
\end{equation}

The stability of the wormhole solution depends upon the conditions
of $V_{eff}^{\prime \prime }\left( \mathcal{R}_{0}\right) >0$ and $%
V_{eff}^{\prime }\left( \mathcal{R}_{0}\right) =V_{eff}\left( \mathcal{R}%
_{0}\right) =0$ as
\begin{equation}
V_{eff} \sim \frac{1}{2}V_{eff}^{\prime \prime }\left( 
\mathcal{R}_{0}\right) \left( \mathcal{R}-\mathcal{R}_{0}\right) ^{2}.
\end{equation}%
Let us then introduce $x=\mathcal{R}-\mathcal{R}_{0}$ and write the equation of
motion again:%
\begin{equation}
\dot{x}^{2}+\frac{1}{2}V_{eff}^{\prime \prime }\left( \mathcal{R}_{0}\right)
x^{2}=0
\end{equation}%
which after a derivative with respect to time reduces to%
\begin{equation}
\ddot{x}+\frac{1}{2}V_{eff}^{\prime \prime }\left( \mathcal{R}_{0}\right)
x=0.
\end{equation}%
Then, using Eqn.(\ref{eq:stress}), we show the conditions for a positive stability value. Our
main aim is to discover the behavior of $V_{eff}^{\prime \prime
}\left( \mathcal{R}_{0}\right) $  as%

\begin{equation}
V_{eff}^{\prime \prime }={\frac {1}{{{\mathcal{R}_{0}}}^{10}} \lbrack -40\, ({\mathcal{R}_{0}}\,\sqrt {{
\frac {{{\mathcal{R}_{0}}}^{4}+2\,{a}^{2}}{{{\mathcal{R}_{0}}}^{4}}}})^{-2\,\omega}{{
\mathcal{R}_{0}}}^{-4\,\omega}{a}^{2}+2\,{{\mathcal{R}_{0}}}^{10}+ \left( 12\,{a}^{2}-12
 \right) {{\mathcal{R}_{0}}}^{6}+40\,{a}^{2}{{\mathcal{R}_{0}}}^{4} \rbrack }. \end{equation}
Note that $a=\omega =0$ corresponds to a non-rotating case.
It is easy to see that $a$ has a crucial role in this stability analysis; the stability regions are  shown in the Figs.(1), (2), (3) and (4). 

\begin{figure}[t]
\begin{center}
\includegraphics[width=10cm]{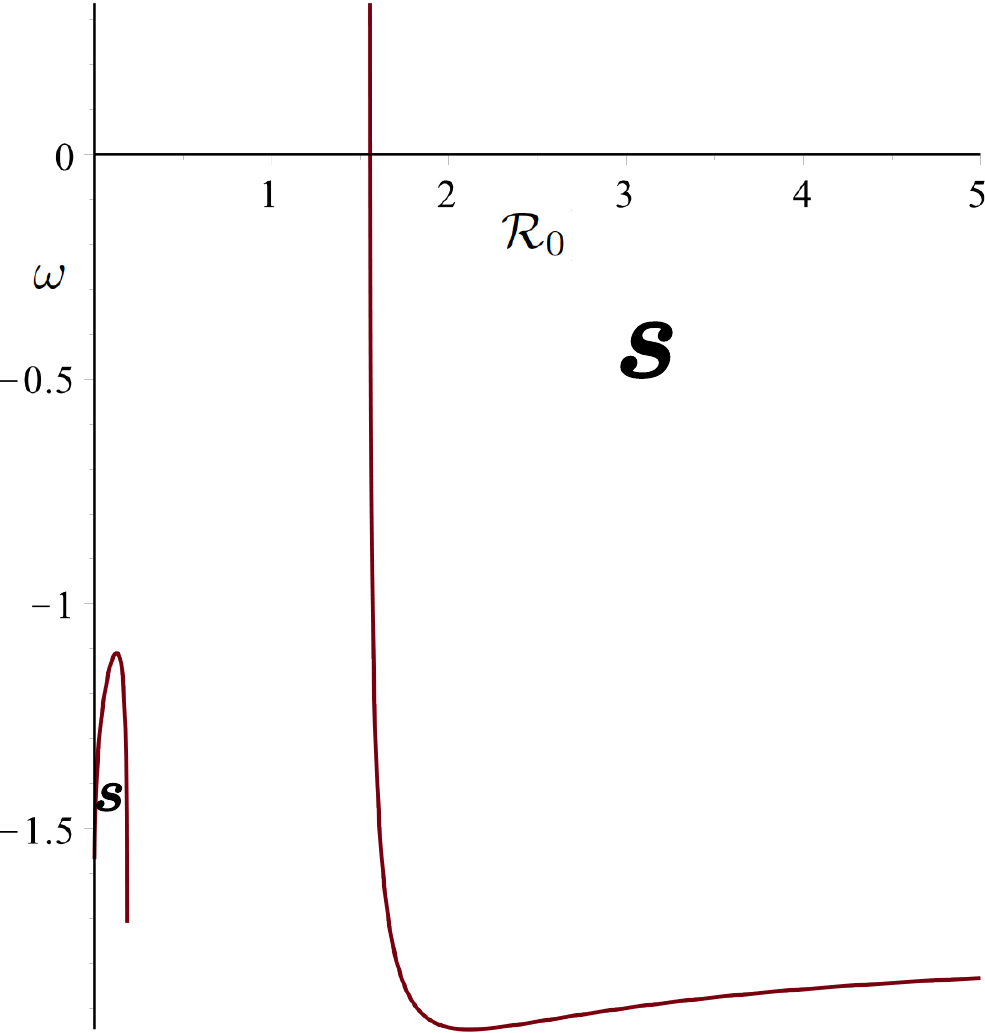}
\end{center}

\caption{Stability of wormhole supported by linear gas in terms of $\protect\omega$ and $%
R_{0}$ for $a$ = 0.1. }
\label{fig:a01}
\end{figure}

\begin{figure}[t]
\begin{center}
\includegraphics[width=10cm]{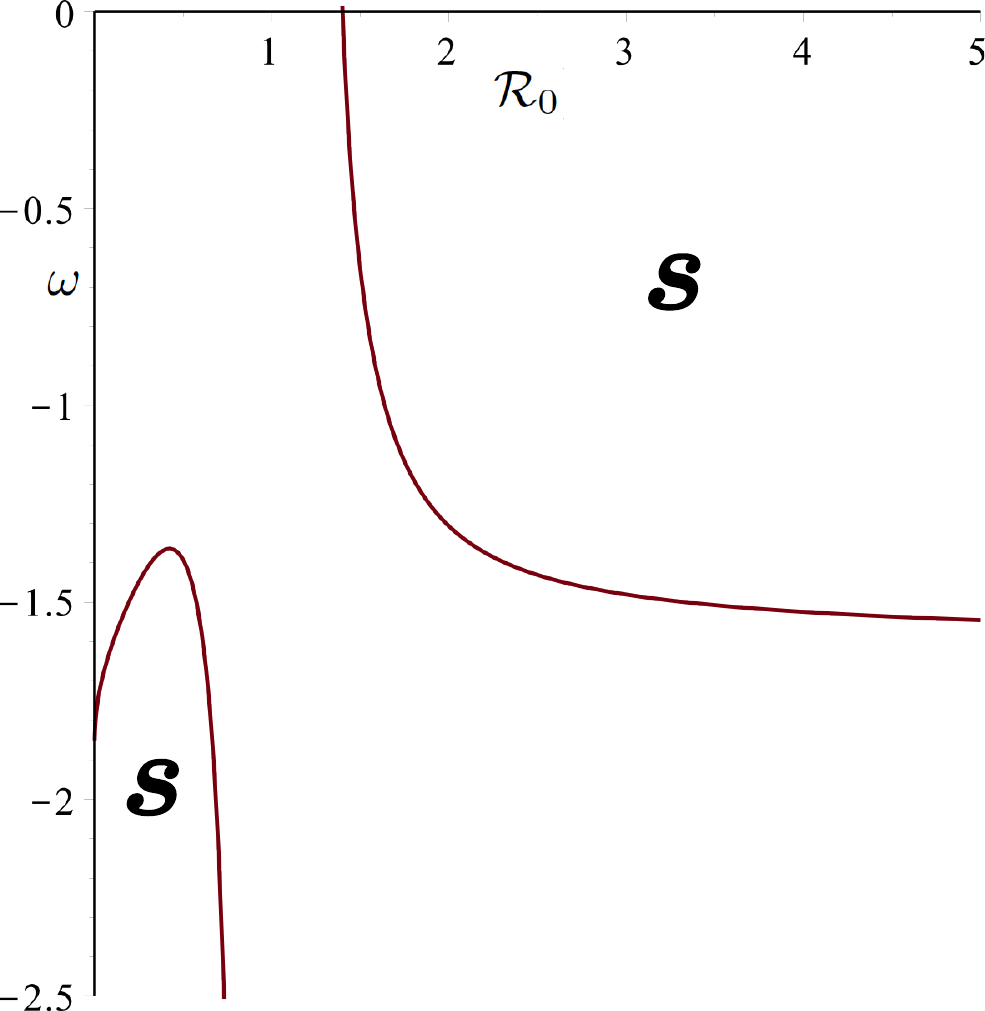}
\end{center}
\caption{Stability of wormhole supported by linear gas  in terms of $\protect\omega$ and $%
R_{0}$ for $a$ = 0.4.}
\label{fig:a10}
\end{figure}

\begin{figure}[t]
\begin{center}
\includegraphics[width=10cm]{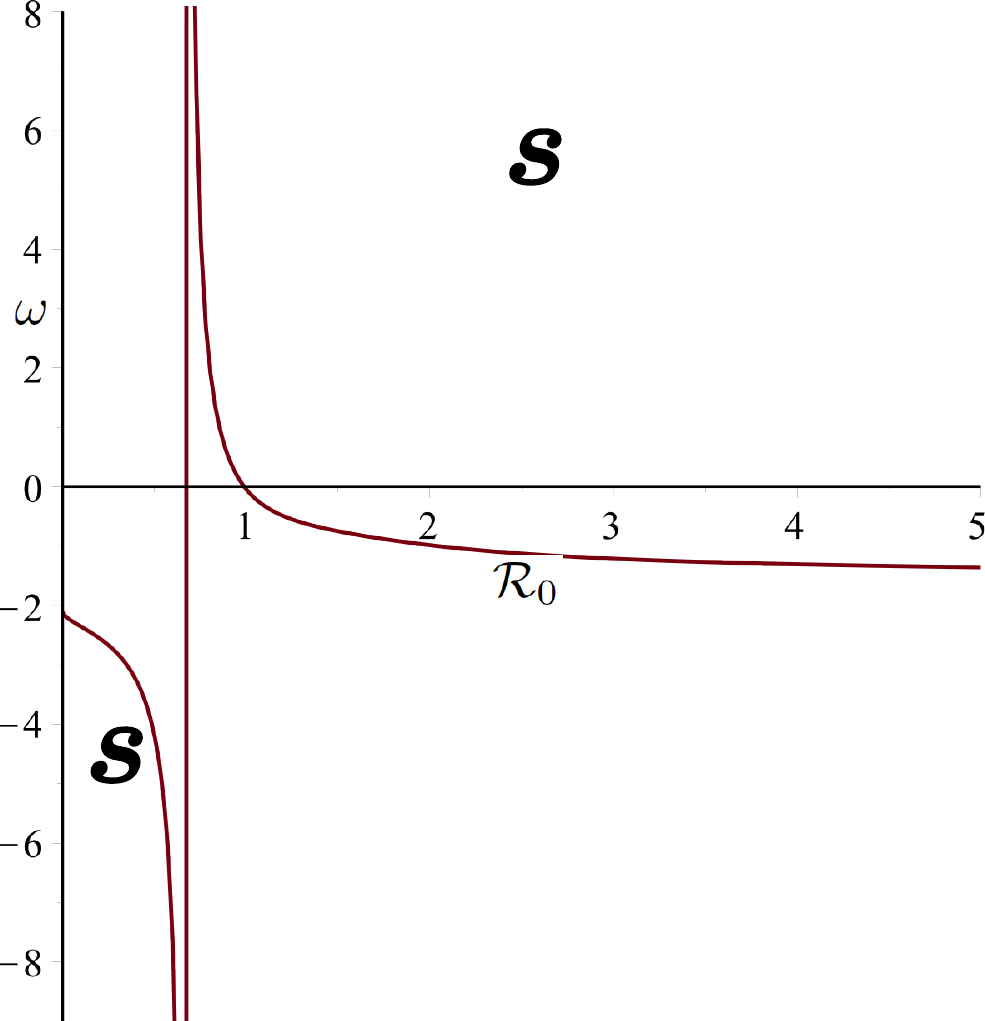}
\end{center}

\caption{Stability of wormhole supported by linear gas  in terms of $\protect\omega$ and $%
R_{0}$ for $a$ = 1.}
\label{fig:a11}
\end{figure}

\begin{figure}[t]
\begin{center}
\includegraphics[width=10cm]{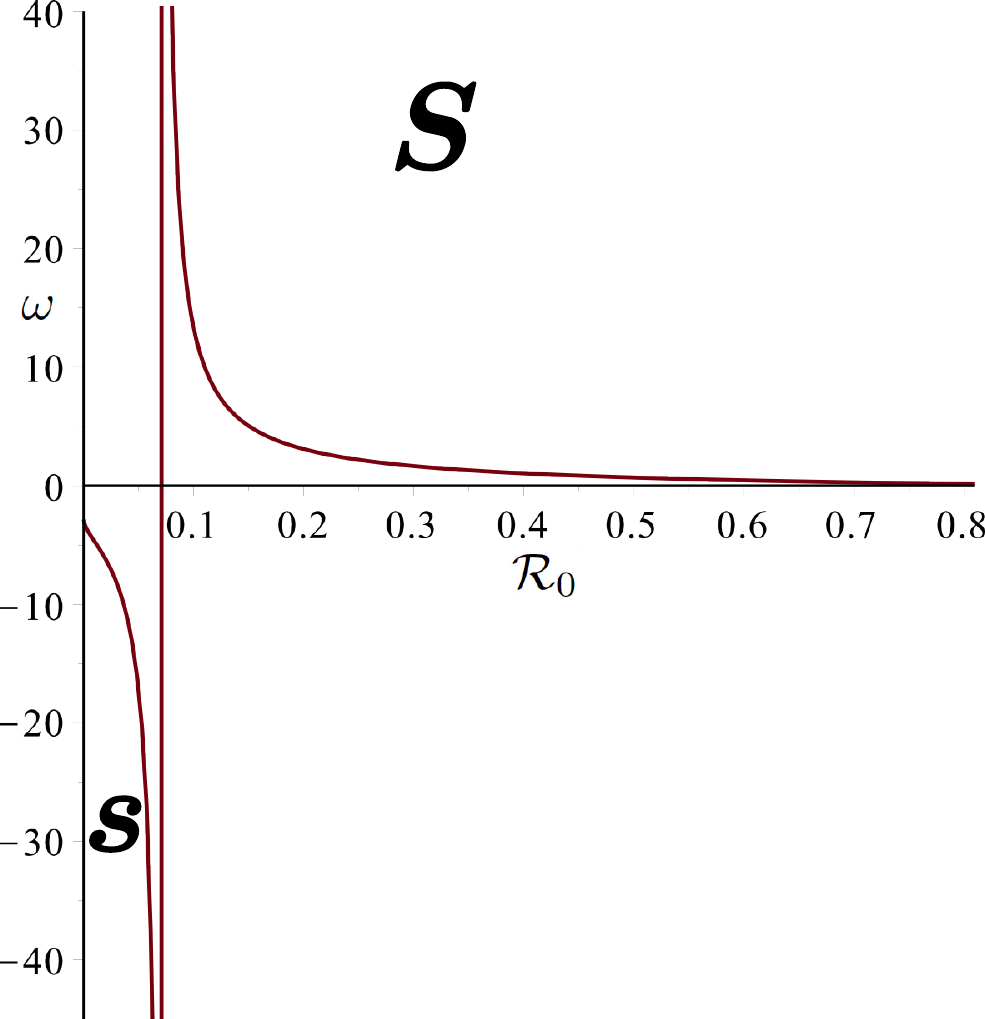}
\end{center}

\caption{Stability of wormhole supported by linear gas  in terms of $\protect\omega$ and $%
R_{0}$ for $a$ = 10.}
\label{fig:a10}
\end{figure}

\section{Discussion}

In this paper, we have studied a thin-shell traversable wormhole with rotation
in five dimensions constructed using a Myers-Perry black hole
with cosmological constants using a cut-and-paste procedure. The standard
stability approach has been applied by considering a linear gas model at the
wormhole throat. Then, the solutions were worked numerically by solving the
dynamical equations and plotted to show the results corresponding
to stability analysis. A key feature of the current analysis is the inclusion of rotation in the
form of non-zero values of angular momentum. Another key aspect of the
current analysis is the focus on different values of parameter ($a$), which plays a crucial role in making the wormhole more stable in five
dimensions. Hence, we observe that the stability of the wormhole is
fundamentally linked to the behavior of the constant ($a$) as shown
in Figs. (1) through (4). The amount of exotic matter required to support the wormhole
is always a crucial issue; unfortunately, we are not able to completely eliminate exotic matter during the constructing of the stable rotating thin-shell wormhole.

\end{document}